# Comparing the diversity of information by word-of-mouth vs. web spread

A. Sela[1(a)], Louis Shekhtman[2], Shlomo Havlin[2], I. Ben-Gal[1]

[1] *Tel Aviv University, Haim Levanon St. 55, 6997801 Tel Aviv Israel*
[2] *Department of Physics, Bar Ilan University* - Ramat Gan 52900, Israel

**Abstract** – Many studies have explored spreading and diffusion through complex networks. The following study examines a specific case of spreading of opinions in modern society through two spreading schemes – defined as being either through 'word-of-mouth' (WOM), or through online search engines (WEB). We apply both modelling and real experimental results and compare the opinions people adopt through an exposure to their friend`s opinions, as opposed to the opinions they adopt when using a search engine based on the PageRank algorithm. A simulated study shows that when members in a population adopt decisions through the use of the WEB scheme, the population ends up with a few dominant views, while other views are barely expressed. In contrast, when members adopt decisions based on the WOM scheme, there is a far more diverse distribution of opinions in that population. The simulative results are further supported by an online experiment which finds that people searching information through a search engine end up with far more homogenous opinions as compared to those asking their friends.

**Introduction.** – Diffusion processes through complex networks have been studied in the context of disease epidemics [1, 2, 3, 4] the spread of computer viruses [4, 5] as well as in the context of information spreading among people [6, 7, 8, 9, 10, 11, 12, 13, 14, 15, 16, 17, 18].

While many of the spreading models are general enough to provide insights on different spreading phenomena, such as detection of influential spreaders, system failures and influence of topologies [4, 19], there are factors that are mainly relevant to information spreading through social networks.

In the context of individuals who adopt opinions, the choice is often among many opinions [20], unlike models for disease spreading [5, 21] or spreading of computer viruses, where a node is either infected or uninfected. Another unique factor to information spreading is that modern information spreading can occur via either physical or virtual interactions. In the process of a virus spreading, an infection tends to occur through the local interactions during a human-human encounter. This resembles information spread by word-of-mouth (WOM), where information diffuses only along the links of the network. Another common method, by which information spreads globally throughout society, is through the internet (WEB) [23]. Such internet interactions are global in their nature and are often mitigated through a search engine.

An example of the type of decisions made through social influence is the choice of where to travel for vacation. In a network where influence occurs only through word-of-mouth, individuals will search for information through their friends about their recent vacations recommendations, and will then decide based upon the different suggestions received from their friends. In contrast, if an individual chooses to use the internet to look for a vacation location, he will probably use a search engine, which will provide him with the requested information.

Several previous works have studied the interactions between word-of-mouth and mass media [22, 23, 24, 25] through Big Data meme tracking methods. Other works came to varying conclusions about the degree to which search engines based on PageRank-related algorithms, [27] bias their search traffic results [26, 13] and amplify the dominance of popular sites. However, none of these works considered the specific comparison, between the spread of ideas through search engines and the spread of ideas through word-of-mouth.

The present work develops an approach for studying modern information diffusion. It considers not only the biases in information flow resulting from the search engines' ranking

(a) alonsela@tauex.tau.ac.il

algorithms, but also the bias which results from human behaviour and tendencies in the context of web searches. Such a bias can only be evaluated through a direct comparison between these two spreading mechanisms.

The spread of opinions by WOM and WEB have much in common. In both cases, a person is influenced by the opinions he has been exposed to and, thus, selects an opinion among these alternatives. In both cases, the social influence [12, 28, 29, 30, 31], will have a significant impact on the person's final decision.

The fundamental difference between the WOM spreading and the WEB spreading is that in the WOM the source of opinions is generated from real acquaintances while in the WEB it is from opinions fetched by an online search engine.

We develop and simulate models for comparing spreading through WEB and WOM. We find that information spread through WOM results in far more diversified opinions of the network's population (see illustrative example in Fig. 1). These results are further strengthened through an experimental study on real human subjects that supports these claims.

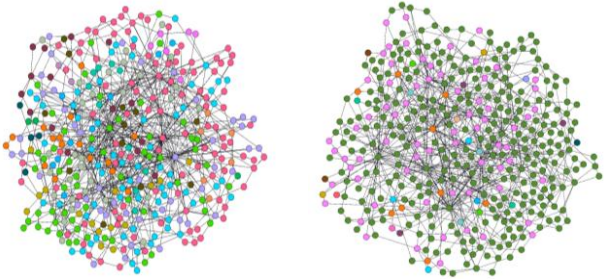

Fig. 1. (Color online) Result of a single realization of 15 opinions' spreading in WOM (left) vs. WEB (right) with a similar selection rule, where each colour represents a different opinion adopted by the node. It can be seen that the WEB spreading results with significantly less diversified opinions in comparison to the WOM spreading.

In the next section, we present the spreading model details, followed by the obtained results from the simulations and the experiments.

**The proposed models.** – The process of information spread can be divided into two stages: (i) an *awareness stage* when a user only becomes aware of a new topic; and (ii) an *evaluation stage*, when a user is exposed to opinions on the topic and has to select which opinion among these alternatives he / she would adopt.

While the awareness stage is similar in both the WEB and the WOM models, the evaluation stage is different. In the awareness stage, for both models, the user first becomes aware of the existence of a new topic by a neighbouring node which holds an opinion on the topic. After the user first becomes aware of a new topic from his neighbour, the user searches information on the new topic through either WOM or through WEB methods, and evaluates the information found in order to form his own personal opinion. In the evaluation stage, users are exposed to different opinions from different sources. In the WOM model they seek opinions from their social connections, e.g., family and friends, while in the WEB model they are exposed to opinions that are presented by the search engines following some online query.

For example, consider a user hearing his work colleagues talking about their locations for their summer vacation (awareness). The user might then seek for information about a location for his own summer vacation. The user might search for such a location by asking his friends for their recommendations (WOM) or he can search for such a location online through a search engine (WEB). The user will then evaluate among the options considered and reach a decision for his / her vacation destination.

Information evaluation via WOM has been the subject of several studies [10, 15, 32, 33, 34]. In these studies, social influence is often modelled by the probability for adopting an opinion, which increases with the number of people holding this opinion in one`s social circle. Similarly, the adoption of an opinion in the WEB is the outcome of similar social and cognitive processes. Thus, in general, the probability for adopting an opinion is proportional to its popularity, whether it is promoted by actual social connections or by web pages. In the WEB, as in WOM, the probability for an adoption of an opinion increases as more web pages support this opinion. Apart from the fact that in the case of the WEB these opinions are collected globally by a search engine, and are written online, similar cognitive evaluation process are performed both for the WOM as for the WEB.

While the detailed algorithm for ranking pages by search engines is not fully known, PageRank is considered to be one of their most important aspects. The PageRank algorithm ranks well connected web pages with higher grades, and the search engine places the links to these highly ranked web pages at the top of the search results list. In our WEB model we define the network as the network of users, i.e. the readers and the publishers of opinions on the internet. We assume that highly connected individuals publish their opinions in highly connected webpages. Accordingly, we calculate the PageRank score of the web page that publishes an opinion by the PageRank score of the person that holds this opinion. This score is then used to set the position of the opinion in the search engine result list.

After the different opinions are ordered according to the PageRank of their publisher, the searchers read these opinions as if they were links in the search query list. It is well known that the higher a search result appears in the search result list, the more likely it will be read by the user. This tendency is expressed through the Search Engine Result Page (*SERP*) function, which defines the probability of a person clicking on a link as a function of the relative position of that link in the search result. The SERP function is a known probabilistic function that has been estimated from several surveys that are mainly performed by search engine optimization (SEO) firms. We estimated the SERP function on the basis of 8 different surveys, which were conducted between years 2006-2014 by 6 different SEO firms as found in [35] and in the firm's web sites.

The following section presents some basic notation and a more formal description of the WEB and WOM spreading dynamics.

The spread dynamics
Let $G = (V, E)$ be a social network of $|V| = N$ participants (nodes). At time $t=0$, a small subset $V` \subseteq V$ of nodes is randomly chosen, and each node is seeded with an opinion from the vector of all possible opinions $B = \{b_1, b_2, ..., b_l\}$. The spreading process then begins, using either (i) WOM or (ii) WEB spreading, such that the opinion held by node $i$ at time $t$ is denoted $o_i^t$. Each user (node), is only able to read a limited number of $k$ opinions from among the existing opinions. This limitation is especially important considering the vast amount of information available in the WEB which can never be fully read. While in the WOM model, different opinions from the node`s social circle are read with a similar probability, in the WEB model, the probability of a node considering an opinion is derived according to the SERP function and the opinion`s position in the search results. More precisely, in the WEB model, a list of all the opinions in the network are first sorted by the PageRank of the node holding the opinion, and then $k$ opinions are chosen to be read from this list as derived from the SERP function. This process continues until all the nodes in the network have adopted an opinion.

Once the spreading process ends, the final adoption fractions of each different opinion in the network are recorded while being sorted (in descending order) in the vector of final adoption fractions $R^{end}$. We note that the relative adoption fractions at late intermediate stages are found to be similar to the final adoption fraction where all the nodes accept an opinion.

After a node has adopted an opinion, a later change of opinion is not permitted in the current model. The rational for not allowing a change of opinion is the cost of opinion change. For example, cancelling a vacation after ticketing, can results in cancelation fees that would prevent (in most cases) such change of vacation destination after the act of conclusion has been made. Thus, the proposed model does not allow a node to alter its opinion once the selection was made.

The next section explicitly defines the spreading process through WEB and WOM schemes.

The WOM spreading process
*While not all nodes infected*
  *For each non-infected node u which has at least one infected neighbour*
  1. Create a list of the influencers opinions held by the neighbours of u that have an opinion (define as IO).
  2. Choose a random opinion from the list IO. Note that opinions present more often among the neighbours are more likely to be chosen.
  3. Adopt the chosen opinion from step 2.
  *Advance time in one time step.*

The WEB spreading process
*While not all nodes infected*
  *For each non-infected node u which has at least one infected neighbour*
  1. Create a list of all the opinions of all the nodes in the network which have any opinion.
  2. Sort the list by the PageRank of the node that holds the opinion, (defined this list as AO).
  3. Create from AO, a second list of k entries (opinions) which represent the actual opinions that would be read by an average user, (denoted IO for Influencers Opinions). In the creation of IO from AO, the probability of reading an opinion located at position i in AO is given by the SERP probability function.
  4. Choose a random opinion from the list IO.
  5. Adopt the chosen opinion from step 4.
  *Advance time in one time step.*

In the next section, we will present the simulation results, followed by results from an experiment with human subjects, which support the simulative results.

**Results**
Simulation results
The simulation set includes 8,100 runs of opinions' spreading under different conditions and parameters, as indicated in Table 1. Overall, in each simulation run, a network of size $N$ was constructed, by implementing a preferential attachment process [4], in which each new node connects to $m$ new nodes. The degree of preferential attachment process, denoted *PA*, varies with *PA*=1 being a fully preferential attachment process, *PA*=0 representing an Erdos-Renyi network, and *PA*=0.5 being a process where in 50% of cases a random node is chosen, and in 50% of cases the selection is governed by a preferential attachment process.

For each combination of the parameters in Table 1, 25 realizations were simulated, summing up in 8100 realizations overall. The vector $R^{end}$ of the final fractions of opinions' spread, for each of the 45 initially seeded ideas was recorded, and sorted in descending order.

As seen in Fig. 2 and Fig. 3, the final fraction of ideas in the network is less diverse in the WEB model than that of the WOM model. For example, in Fig. 2, the most common idea was adopted by approximately 75% of the nodes in the WEB spread, but only 23% of the nodes in the WOM spread. Furthermore, in the WEB spread over 95% of the population adopted on average only three top ideas, while in the WOM spread 95% of the population adopted as much as 15 ideas, and each of them was adopted by a relatively large fraction of the population.

The distribution of values in the adoption fraction vector $R^{end}$ for the 8,100 runs, for the first 6 most common ideas, as seen in Fig. 3, reveals a right peak in the WEB spread histogram for the 1st idea, between the population fraction of 0.85 and 1.

A. Sela *et al.*

Table 1: **Summary of simulation parameters**

| Denote | Parameter | Possible Values |
|---|---|---|
| PA | Preferential attachment level | 0, 0.5, 1 |
| N | Number of nodes | 5000, 10000[1] |
| m | Network density | 2, 4, 8 |
| k | Number of opinions read | 5, 10, 15 |
| II | Number of nodes infected at $t=0$. | 15, 30, 45 |
| S | Spreading model | WOM, WEB |

This peak is the outcome of simulations runs where one single idea is adopted by a large fraction of nodes in the network. The dominance of one single idea in the WEB spread can be seen more clearly (see Fig. 3) when comparing the fraction of the population which adopted the 1$^{st}$ most popular idea, 2$^{nd}$ most popular, 3$^{ed}$ most popular idea etc. (see Fig. 3).

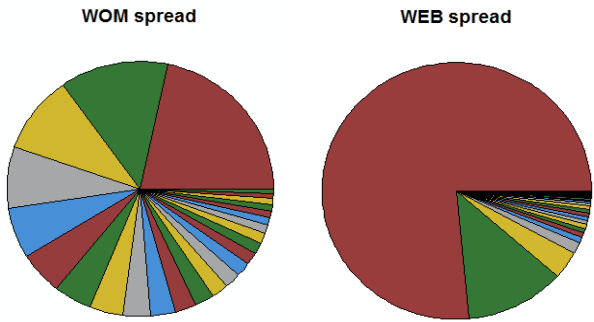

Fig. 2. (Colour online) Final average fractions of adoption for spread of different opinions, as generated by the WOM and the WEB simulations when using Table 1 parameters. Note that the WOM model results in a significant higher variability of opinions' spread in the population.

The adoption fraction of the 1$^{st}$ most popular idea for the WEB spreading model (red histogram) is significantly larger on average than that for the WOM spreading model, where the 1$^{st}$ idea in the WEB spreading follows a wide distribution with adoption fractions varying between 0.3-0.99. In comparison, the 1$^{st}$ idea in WOM spreading model (azure histogram), has a lower mean adoption rate of approximately 0.23, and follows a narrower Gaussian distribution. This trend flips, from the 2$^{nd}$ most popular idea onward, where the mean of the WOM model is larger than the mean of the WEB model. When comparing the adoption fractions in the 7$^{th}$, 8$^{th}$ and 9$^{th}$ popular ideas, in the WOM model these ideas still capture a reasonable fraction of the population, whereas in the WEB model these ideas have barely spread.

Experimental Results with Human Subjects

To test our conclusion that the use of the WEB method results in more homogenous opinions in a population, we conducted an experiment based on real human subjects that were asked to use either the WOM information search approach or the WEB approach.

Two groups of users were required to answer the same set of questions. One group was requested to answer the question solely by using the Google search engine, while the other was instructed to answer the questions by asking their friends and was instructed not to use any search engine. The three questions were:

1. *What is the best new car to buy?*
2. *What is the best country for a vacation overseas?*
3. *What is the best restaurant in New York?*

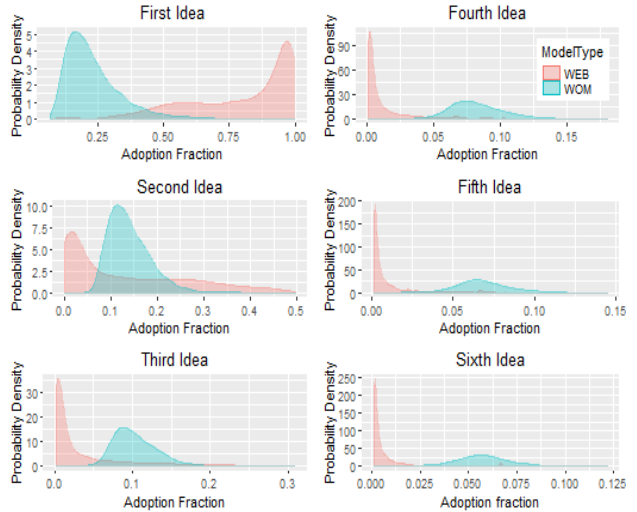

Fig. 3. Adoption fractions of six top ideas in the network, sorted by their popularity (1$^{st}$ idea is the most common one) for both the WOM (marked by azure) and the WEB (marked by red) spreads. Note that while the most common idea spreads to a large fraction of the network by the WEB, the less common ideas can be observed for the WOM spread but are barely noticeable in WEB spread.

These three questions were answered by 100 Mechanical Turk responders, of which 50 used the WEB model and 50 used by the WOM model. After cleaning the data and combining similar answers such as "London" and "England" in Question 1, the final results included 49 WEB responders and 49 WOM responders, each of whom answered all three questions and a total of 294 complete answers have been reported.

Fig. 4 shows that the WEB spreading results in a fewer ideas being adopted by a larger number of responders. For example, UK was repeatedly indicated as the best location for vacation in 26 out of 49 responses (53%) among the WEB users, while Australia and Japan were most popular in the WOM model with only 6 out of 49 users (12%). Furthermore, as can be seen in Table 2, while the WEB model resulted in 17 different opinions for the "best restaurant" question, and as much as 16 responders repeating the same name of restaurant to be the best restaurant in NY, the WOM model included as many as 38 different "best restaurant" answers with only 4 repeated names of restaurants, thus, the experimental results strongly support the model simulation results. While all questions included a lower variability of information while using the WEB as compared to the WOM, the most extreme reduction in the diversity of information is seen in NY restaurants question as presented in Fig. 4 and Fig. 5.

---

[1] Several specific runs with networks of sizes N=20,000 and 30,000 were also inspected in order to verify that a larger network size is consistent with the simulation results. These results were not incorporated in the entire simulations analysis due to their long running times by Agent Based Modelling (ABM) simulation methodology.

Table 2 – Answers to 3 questions

|  | WEB | WOM |
|---|---|---|
| **Question 1 – Best car** | | |
| Number of different uniqe answers | 24 | 43 |
| Number of repetitons for the most common answer | 12 | 3 |
| **Question 2 – Country for vacation** | | |
| Number of different uniqe answers | 16 | 23 |
| Number of repetitons for the most common answer | 26 | 6 |
| **Question 3 – Best restaurant in NY** | | |
| Number of different uniqe answers | 17 | 38 |
| Number of repetitons for the most common answer | 16 | 4 |

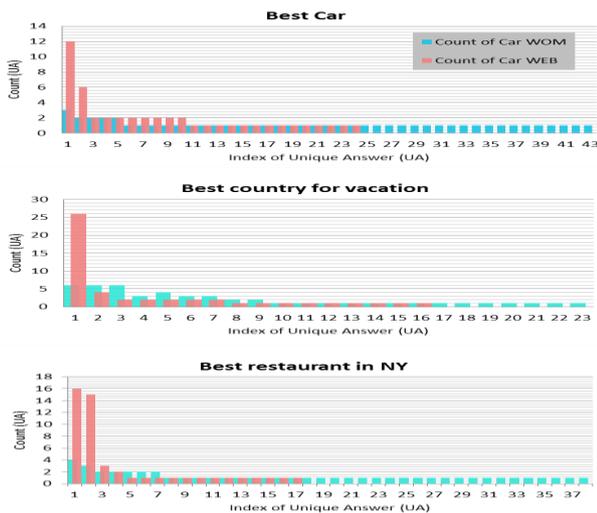

Fig. 4. The distribution of answers for the three questions ("best car", "best country for vacation", "best restaurant in NY") obtained by the WOM search (azure) vs. the WEB search (red). It can be clearly seen that the WEB search results in more similar answers among the population while the WOM search results in more diverse answers.

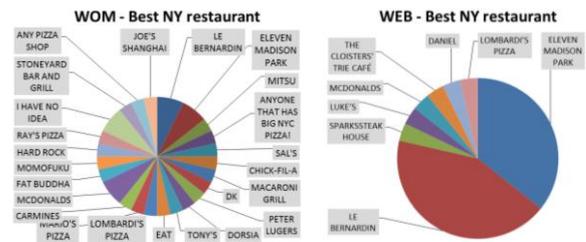

Fig. 5. Information diversity for the question "best restaurant in NY" as received from users using WOM vs. WEB searching methods

**Conclusion -** Our results suggest that the use of WEB search engines substantially decreases the diversity of opinions in a population, compared to word-of-mouth (WOM) spreading. While previous studies have attempted to suggest that web search results are less biased than believed [27] and that the distribution of internet pages is less unbalanced than expected, we suggest that users' decisions are still highly biased when using the WEB search engine since each user ends up reading similar opinions for similar searches. This is the result of two independent "rich get richer" processes, where the first is in the search engine algorithm and the second is in users' behaviour as expressed in the SERP function. Such similarity in the exposure to opinions might lead users to make similar decisions and thus increases homogeneity in the population.

The importance of diversity is well known in many scientific fields, including the key role of a genetic diversity as a way for populations to adapt to changing environments. Diversity of opinions is also known to have its advantages in creative processes [36]. In cases where a diversity of opinions is required, this work recommends to include (at least to some degree) the WOM information search and spread, which can be obtained by attending conferences or using social networks which are seen as WOM information search. These recommendations are particularly important as people rely more and more on search engines. Measuring the influence on creative processes when solely using search engines as a tool for information search can be a subject for further future research.